# Nutation Resonance in Ferromagnets

Mikhail Cherkasskii[1,*], Michael Farle[2,3], and Anna Semisalova[2]

[1] *Department of General Physics 1, St. Petersburg State University, St. Petersburg, 199034, Russia*
[2] *Faculty of Physics and Center of Nanointegration (CENIDE), University of Duisburg-Essen, Duisburg, 47057, Germany*
[3] *Kirensky Institute of Physics, Federal Research Center KSC SB RAS, Russia*

* Corresponding author: macherkasskii@hotmail.com

The inertial dynamics of magnetization in a ferromagnet is investigated theoretically. The analytically derived dynamic response upon microwave excitation shows two peaks: ferromagnetic and nutation resonances. The exact analytical expressions of frequency and linewidth of the magnetic nutation resonance are deduced from the frequency dependent susceptibility determined by the inertial Landau-Lifshitz-Gilbert equation. The study shows that the dependence of nutation linewidth on the Gilbert precession damping has a minimum, which becomes more expressive with increase of the applied magnetic field.

PACS numbers: 76.50.+g, 78.47.jp, 75.50.-y

## I. INTRODUCTION

Recently, the effects of inertia in the spin dynamics of ferromagnets were reported to cause nutation resonance [1-12] at frequencies higher than the conventional ferromagnetic resonance. It was shown that inertia is responsible for the nutation, and that this type of motion should be considered together with magnetization precession in the applied magnetic field. Nutation in ferromagnets was confirmed experimentally only recently [2], since nutation and precession operate at substantially different time scales, and conventional microwave ferromagnetic resonance (FMR) spectroscopy techniques do not easily reach the high-frequency (sub-Terahertz) regime required to observe the inertia effect which in addition yields a much weaker signal.

Similar to any other oscillatory system, the magnetization in a ferromagnet has resonant frequencies usually studied by ferromagnetic resonance [13,14]. The resonant eigenfrequency is determined by the magnetic parameters of the material and applied magnetic field. Including inertia of the magnetization in the model description shows that nutation and precession are complementary to each other and several resonances can be generated. In this Letter, we concentrate on the investigation of the resonance characteristics of nutation.

The investigation of nutation is connected to the progress made in studies of the spin dynamics at ultrashort time scales [15,16]. These successes led to the rapid development of a new scientific field, the so-called ultrafast magnetism [17-25]. The experimental as well as theoretical investigation of the inertial spin dynamics is at the very beginning, although it might be of significance for future high speed spintronics applications including ultrafast magnetic switching.

Besides nutation driven by magnetization inertia, several other origins of nutation have been reported. Transient nutations (Rabi oscillations) have been widely investigated in nuclear magnetic resonance [26] and electron spin resonance [27-29], they were recently addressed in ferromagnets [30]. A complex dynamics and Josephson nutation of a local spin $s = 1/2$ as well as large spin cluster embedded in the tunnel junction between ferromagnetic leads was shown to occur due to a coupling to Josephson current [31-33]. Low-frequency nutation was observed in nanomagnets exhibiting a non-linear FMR with the large-angle precession of magnetization where the onset of spin wave instabilities can be delayed due to geometric confinement [34]. Nutation dynamics due to inertia of magnetization in ferromagnetic thin films was observed for the first time by Neeraj et al. [2].

The microscopic derivation of the magnetization inertia was performed in ref. [3-7]. A relation between the Gilbert damping constant and the inertial regime characteristic time was elaborated in ref. [3]. The exchange interaction, damping, and moment of inertia can be calculated from first principles as shown in [7]. The study of inertia spin dynamics with a quantum approach in metallic ferromagnets was performed in [8]. In addition, nutation was theoretically analyzed as a part of magnetization dynamics in ferromagnetic nanostructure [9,10] and nanoparticles [11]. Despite these advances, exact analytical expressions for the high-frequency susceptibility including inertia had not been derived yet.

In [35], the inertial regime was introduced in the framework of the mesoscopic nonequilibrium thermodynamics theory, and it was shown to be responsible for the nutation superimposed on the precession of magnetization. Wegrowe and Ciornei [1] discussed the equivalence between the inertia in the dynamics of uniform precession and a spinning top within the framework of the Landau–Lifshitz–Gilbert equation generalized to the inertial regime. This equation was studied analytically and numerically [12,36]. Although these reports provide numerical tools for obtaining resonance characteristics, the complexity of the numerical solution of differential equations did not allow to estimate the nutation frequency and linewidth accurately. Also in a recent remarkable paper [37] a novel collective excitation – the nutation wave – was reported, and the dispersion characteristics were derived without discussion of the nutation resonance lineshapes and intensities.

Thus, at present, there is a necessity to study the resonance properties of nutation in ferromagnets, and this paper is devoted to this study. We performed the investigation based on the Landau-Lifshitz-Gilbert equation with the additional inertia term and provide an analytical solution.

It is well known that the Landau-Lifshitz-Gilbert equation allows finding the susceptibility as the ratio between the time-varying magnetization and the time-varying driving magnetic field (see for example [38,39] and references therein). This susceptibility describes well the magnetic response of a ferromagnet in the linear regime, that is a small cone angle of the precession. In this description, the ferromagnet usually is placed in a magnetic field big enough to align all atomic magnetic moments along the field, i.e., the ferromagnet is in the saturated state and the magnetization precesses. The applied driving magnetic field allows one to observe FMR as soon as the driving field frequency coincides with eigenfrequency of precession. Using the expression for susceptibility, one can elaborate such resonance characteristics as eigenfrequency and linewidth. We will present similar expressions for the dynamic susceptibility, taking nutation into account.

## II. SUSCEPTIBILITY

The ferromagnet is subjected to a uniform bias magnetic field $\mathbf{H}_0$ acting along the z-axis and being strong enough to initiate the magnetic saturation state. The small time-varying magnetic field $\mathbf{h}$ is superimposed on the bias field. The coupling between impact and response, taking into account precession, damping, and nutation, is given by the Inertial Landau-Lifshitz-Gilbert (ILLG) equation

$$\frac{d\mathbf{M}}{dt} = -|\gamma|\mathbf{M} \times \left[\mathbf{H}_{\mathit{eff}} - \frac{\alpha}{|\gamma|M_0}\left(\frac{d\mathbf{M}}{dt} + \tau \frac{d^2\mathbf{M}}{dt^2}\right)\right], \quad (1)$$

where $\gamma$ is the gyromagnetic ratio, $\mathbf{M}$ the magnetization vector, $M_0$ the magnetization at saturation, $\mathbf{H}_{\mathit{eff}}$ the vector sum of all magnetic fields, external and internal, acting upon the magnetization, $\alpha$ the Gilbert damping, and $\tau$ the inertial relaxation time. For simplicity, we assume that the ferromagnet is infinite, i.e. there is no demagnetization correction, with negligible magnetocrystalline anisotropy, and only the externally applied fields contribute to the total field. Thus, the bias magnetic field $\mathbf{H}_0$ and signal field $\mathbf{h}$ are included in $\mathbf{H}_{\mathit{eff}}$. We assume that the signal is small $|\mathbf{h}| \ll |\mathbf{H}_0|$, hence the magnetization is directed along $\mathbf{H}_0$.

Our interest is to study the correlated dynamics of nutation and precession simultaneously; therefore we write the magnetization and magnetic field in the generalized form using the Fourier transformation

$$\mathbf{M}(t) = M_0 \hat{z} + \frac{1}{\sqrt{2\pi}} \int_{-\infty}^{\infty} d\omega' \, \mathbf{m}(\omega') e^{i\omega't}, \quad (2)$$

$$\mathbf{H}_{\mathit{eff}}(t) = H_0 \hat{z} + \frac{1}{\sqrt{2\pi}} \int_{-\infty}^{\infty} d\omega' \, \mathbf{h}(\omega') e^{i\omega't}, \quad (3)$$

where $\hat{z}$ is the unit vector along the z-axis. If we substitute these expressions in the ILLG equation and neglect the small terms, it leads to

$$\frac{1}{\sqrt{2\pi}} \int_{-\infty}^{\infty} d\omega' \, i\omega' \mathbf{m}(\omega') e^{i\omega't} = \frac{1}{\sqrt{2\pi}} \int_{-\infty}^{\infty} d\omega' e^{i\omega't}$$
$$\times \left[-|\gamma| M_0 \hat{z} \times \mathbf{h}(\omega') + |\gamma| H_0 \hat{z} \times \mathbf{m}(\omega') \right. \quad (4)$$
$$\left. + i\alpha \omega' \hat{z} \times \mathbf{m}(\omega') - \alpha \tau {\omega'}^2 \hat{z} \times \mathbf{m}(\omega')\right].$$

By performing the Fourier transform and changing the order of integration, equation (4) becomes

$$\frac{1}{2\pi} \int_{-\infty}^{\infty} d\omega' \int_{-\infty}^{\infty} dt \, i\omega' \mathbf{m}(\omega') e^{i(\omega'-\omega)t}$$
$$= \frac{1}{2\pi} \int_{-\infty}^{\infty} d\omega' \int_{-\infty}^{\infty} dt \, e^{i(\omega'-\omega)t} \quad (5)$$
$$\times \left[-|\gamma| M_0 \hat{z} \times \mathbf{h}(\omega') + |\gamma| H_0 \hat{z} \times \mathbf{m}(\omega') \right.$$
$$\left. + i\alpha \omega' \hat{z} \times \mathbf{m}(\omega') - \alpha \tau (\omega')^2 \hat{z} \times \mathbf{m}(\omega')\right],$$

where the integral representation of the Dirac delta function can be found. With the delta function, the equation (5) simplifies to

$$i\omega \mathbf{m}(\omega) = -|\gamma| M_0 \hat{z} \times \mathbf{h}(\omega) + |\gamma| H_0 \hat{z} \times \mathbf{m}(\omega)$$
$$+ i\alpha \omega \hat{z} \times \mathbf{m}(\omega) - \alpha \tau \omega^2 \hat{z} \times \mathbf{m}(\omega). \quad (6)$$

By projecting to Cartesian coordinates and introducing the circular variables for positive and negative circular polarization $m_\pm = m_x \pm i m_y$, $h_\pm = h_x \pm i h_y$, one obtains

$$-\alpha\tau m_+ \omega^2 + (-m_+ + i\alpha m_+)\omega + (\omega_H m_+ - \omega_M h_+) = 0,$$
$$-\alpha\tau m_- \omega^2 + (m_- + i\alpha m_-)\omega + (\omega_H m_- - \omega_M h_-) = 0, \quad (7)$$

where $\omega_H = |\gamma| H_0$ is the precession frequency and $\omega_M = |\gamma| M_0$. The small-signal susceptibility follows from these equations:

$$m_\pm = \chi_\pm h_\pm,$$

$$\chi_+ = \frac{\omega_M}{\omega_H - \omega - \alpha\tau\omega^2 + i\alpha\omega}, \quad (8)$$

$$\chi_- = \frac{\omega_M}{\omega_H + \omega - \alpha\tau\omega^2 + i\alpha\omega}.$$

It is seen that the susceptibility (8) is identical with the susceptibility for LLG equation, if one drops the inertial term, that is $\tau = 0$.

Let us separate dispersive and dissipative parts of the susceptibility $\chi_\pm = \chi'_\pm - i\chi''_\pm$,

$$\chi'_+ = \frac{-\omega_M(\omega - \omega_H + \alpha\tau\omega^2)}{D_+},$$

$$\chi''_+ = \frac{\alpha\omega\omega_M}{D_+},$$

$$\chi'_- = \frac{\omega_M(\omega + \omega_H - \alpha\tau\omega^2)}{D_-}, \quad (9)$$

$$\chi''_- = \frac{\alpha\omega\omega_M}{D_-},$$

$$D_+ = \alpha^2\tau^2\omega^4 + 2\alpha\tau\omega^3 + (\alpha^2 - 2\alpha\tau\omega_H + 1)\omega^2 - 2\omega_H\omega + \omega_H^2, \quad (10)$$

$$D_- = \alpha^2\tau^2\omega^4 - 2\alpha\tau\omega^3 + (\alpha^2 - 2\alpha\tau\omega_H + 1)\omega^2 + 2\omega_H\omega + \omega_H^2. \quad (11)$$

The frequency dependence of the dissipative parts of susceptibilities $\chi''_+$ and $\chi''_-$ is shown in the Fig. 1. The plus and minus subscripts correspond to right-hand and left-hand direction of rotation. Since the denominators $D_+$ and $D_-$ are quartic polynomials, four critical points for either $\chi''_+$ or $\chi''_-$ can be expected. Two of them that are extrema with a clear physical meaning are plotted. In Fig. 1(a) the extremum, corresponding to FMR at $\omega'_H \approx |\gamma|H_0$ is shown. Due to the contribution of nutation, the frequency and linewidth of this resonance are slightly different from the ones of usual FMR. The resonance occurs for right-hand precession, i.e. positive polarization.

In Fig. 1(b) the nutation resonance possessing negative polarization is presented. Note that the polarizations of ferromagnetic and nutation resonances are reversed.

## III. APPROXIMATION FOR NUTATION FREQUENCY

Let us turn to the description of an approximation of the nutation resonance frequency. If we equate the denominator $D_-$ to zero, solve the resulting equation, we obtain the approximation from the real part of the roots. This is reasonable, since the numerator of $\chi''_-$ is the linear function of $\omega$, and we are interested in $\omega \gg 1$. Indeed, the equation

$$\alpha^2\tau^2\omega^4 - 2\alpha\tau\omega^3 + (\alpha^2 - 2\alpha\tau\omega_H + 1)\omega^2 + 2\omega_H\omega + \omega_H^2 = 0 \quad (12)$$

has four roots that are complex conjugate in pairs

$$w_{-FMR_{1,2}} = \frac{1 \pm i\alpha - \sqrt{1 - \alpha^2 + 4\alpha\tau\omega_H \pm 2i\alpha}}{2\alpha\tau}, \quad (13)$$

$$w_{N_{1,2}} = \frac{1 \pm i\alpha + \sqrt{1 - \alpha^2 + 4\alpha\tau\omega_H \pm 2i\alpha}}{2\alpha\tau}. \quad (14)$$

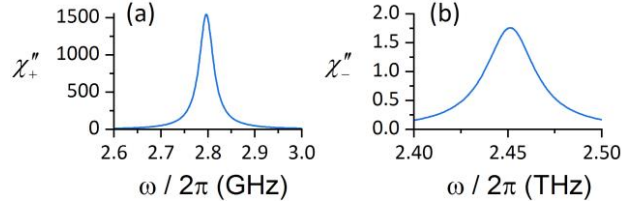

FIG. 1. (Color online) (a) The FMR peak with nutation. (b) The nutation resonance. The calculation was performed for $|\gamma|/(2\pi) = 28$ GHz T$^{-1}$, $\mu_0 M_0 = 1$ T, $\mu_0 H_0 = 100$ mT, $\alpha = 0.0065$ and $\tau = 10^{-11}$ s.

One should choose the same sign from the $\pm$ symbol in each formula, simultaneously. The real part of expression (13) gives the approximate frequency for FMR, but in negative numbers, so the sign should be inversed. The approximate frequency of FMR in positive numbers can be derived from equation $D_+ = 0$. The approximate nutation frequency is obtained by the real part of the expression (14). One takes half the sum of two conjugate roots $w_{N_{1,2}}$, neglects the high $\tau$ terms, and obtains the nutation resonance frequency

$$w_N = \frac{1 + \sqrt{1 + 2\alpha\tau|\gamma|H}}{2\alpha\tau}. \quad (15)$$

Note that the expression of $w_N$ is close to the approximation given in [36] at $\tau \ll 1/\alpha|\gamma|H$, namely

$$\omega_{nu}^{weak} = \frac{\sqrt{1 + \alpha\tau|\gamma|H}}{\alpha\tau}. \quad (16)$$

The similarity of both approximations becomes clear, if we perform a Taylor series expansion and return to the notation $\omega_H$,

$$w_N = \frac{1 + \sqrt{1 + 2\alpha\tau\omega_H}}{2\alpha\tau} = \frac{1}{\alpha\tau} + \frac{\omega_H}{2} - \frac{\alpha\tau\omega_H^2}{4}$$
$$+ \frac{1}{4}\alpha^2\tau^2\omega_H^3 + O(\alpha^3\tau^3),$$

$$\omega_{nu}^{weak} = \frac{\sqrt{1 + \alpha\tau\omega_H}}{\alpha\tau} = \frac{1}{\alpha\tau} + \frac{\omega_H}{2} - \frac{\alpha\tau\omega_H^2}{8}$$
$$+ \frac{1}{16}\alpha^2\tau^2\omega_H^3 + O(\alpha^3\tau^3).$$

## IV. PRECISE EXPRESSIONS FOR FREQUENCY AND LINEWIDTH OF NUTATION

The analytical approach proposed in this Letter yields precise values of the frequency of nutation resonance and the full width at half maximum (FWHM) of the peak. The frequency is found by extremum, when the derivative of the dissipative part of susceptibilities (9) is zero

$$\frac{\partial \chi_-''}{\partial \omega} = 0. \quad (17)$$

It is enough to determine zeros of the numerator of the derivative, that are given by

$$3\alpha^2\tau^2\omega^4 - 4\alpha\tau\omega^3 + \left(\alpha^2 - 2\alpha\tau\omega_H + 1\right)\omega^2 - \omega_H^2 = 0. \quad (18)$$

Let us use Ferrari's solution for this quartic equation and introduce the notation:

$$\begin{aligned}
A_r &= 3\alpha^2\tau^2, \\
B_r &= -4\alpha\tau, \\
C_r &= \alpha^2 - 2\alpha\tau\omega_H + 1, \\
E_r &= -\omega_H^2, \\
a_r &= \frac{C_r}{A_r} - \frac{3B_r^2}{8A_r^2}, \\
b_r &= -\frac{B_r C_r}{2A_r^2} + \frac{B_r^3}{8A_r^3}, \\
c_r &= \frac{B_r^2 C_r}{16A_r^3} - \frac{3B_r^4}{256A_r^4} + \frac{E_r}{A_r}.
\end{aligned} \quad (19)$$

In Ferrari's method, one should determine a root of the nested depressed cubic equation. In the investigated case, the root is written

$$y_r = -\frac{5a_r}{6} + U_r + V_r, \quad (20)$$

where

$$\begin{aligned}
U_r &= \sqrt[3]{-\sqrt{\frac{P_r^3}{27} + \frac{Q_r^2}{4}} - \frac{Q_r}{2}}, \\
V_r &= -\frac{P_r}{3U_r}, \\
P_r &= -\frac{a_r^2}{12} - c_r, \\
Q_r &= \frac{1}{3}a_r c_r - \frac{a_r^3}{108} - \frac{b_r^2}{8}.
\end{aligned} \quad (21)$$

Thus, the precise value of the nutation frequency is given by

$$\Omega_N = -\frac{B_r}{4A_r} + \frac{\sqrt{a_r + 2y_r}}{2} \\ + \frac{1}{2}\sqrt{-3a_r - 2y_r - \frac{2b_r}{\sqrt{a_r + 2y_r}}}. \quad (22)$$

The performed analysis shows that approximate value of nutation resonance frequency is close to precise value.

The linewidth of the nutation resonance is found at a half peak height. If one denotes the maximum by $X_-'' = \chi_-''(\omega = \Omega_N)$, the equation which determines frequencies at half magnitude is

$$\frac{1}{2}X_-''\left[\alpha^2\tau^2\omega^4 - 2\alpha\tau\omega^3 + \left(\alpha^2 - 2\alpha\tau\omega_H + 1\right)\omega^2 \\ + 2\omega\omega_H + \omega_H^2\right] - \alpha\omega\omega_M = 0. \quad (23)$$

We repeat the procedure for finding solutions with Ferrari's method introducing the new notations

$$\begin{aligned}
A_{lw} &= \frac{1}{2}\alpha^2\tau^2 X_-'', \\
B_{lw} &= -\alpha\tau X_-'', \\
C_{lw} &= \frac{1}{2}X_-''\left(\alpha^2 - 2\alpha\tau\omega_H + 1\right), \\
D_{lw} &= \omega_H X_-'' - \alpha\omega_M \\
E_{lw} &= \frac{1}{2}\omega_H^2 X_-'', \\
a_{lw} &= \frac{C_{lw}}{A_{lw}} - \frac{3B_{lw}^2}{8A_{lw}^2}, \\
b_{lw} &= -\frac{B_{lw}C_{lw}}{2A_{lw}^2} + \frac{B_{lw}^3}{8A_{lw}^3} + \frac{D_{lw}}{A_{lw}}, \\
c_{lw} &= \frac{B_{lw}^2 C_{lw}}{16A_{lw}^3} - \frac{3B_{lw}^4}{256A_{lw}^4} - \frac{B_{lw}D_{lw}}{4A_{lw}^2} + \frac{E_{lw}}{A_{lw}}.
\end{aligned} \quad (24)$$

A root of the nested depressed cubic equation $y_{lw}$ must be found in the same way as provided in (20) with the corresponding replacement of variables, i.e. subscript $r$ is replaced by $lw$. The difference between two adjacent roots gives the nutation linewidth

$$\Delta\Omega_N = \sqrt{-3a_{lw} - 2y_{lw} - \frac{2b_{lw}}{\sqrt{a_{lw} + 2y_{lw}}}}. \quad (25)$$

The explicit expression for the linewidth can be written using the equations (19)-(25).

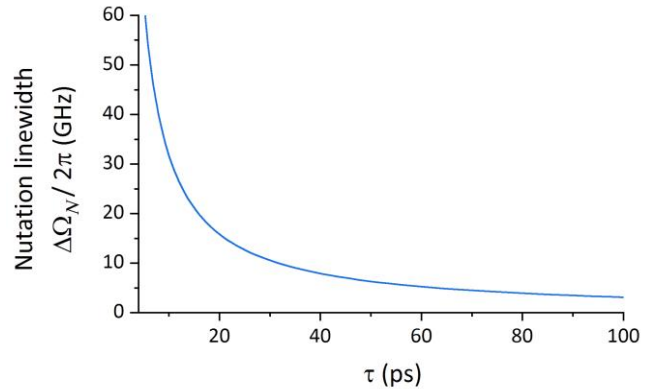

FIG. 2. (Color online) The dependence of the nutation linewidth on the inertial relaxation time for $\mu_0 H_0 = 100$ mT, $\mu_0 M_0 = 1$ T, and $\alpha = 0.0065$.

The effect of the inertial relaxation time on the nutation linewidth is shown in Fig. 2. One can see that increasing inertial relaxation time leads to narrowing of the linewidth. This behavior is expected and is consistent with the traditional view that decreasing of losses results in narrowing of linewidth.

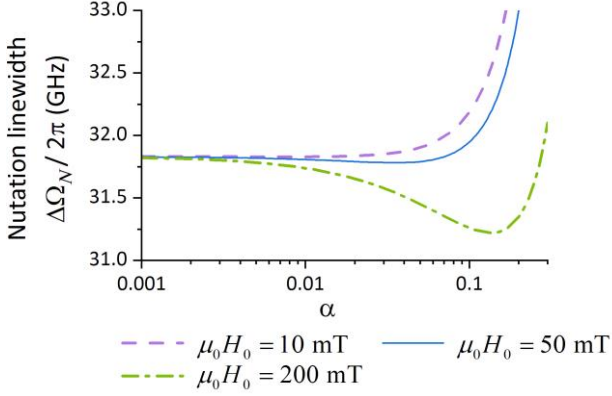

FIG. 3. (Color online) The dependence of nutation resonance linewidth on precession Gilbert damping parameter at different magnetic fields $H_0$ for $\mu_0 M_0 = 1$ T and $\tau = 10^{-11}$ s.

Since the investigated oscillatory system implements simultaneous two types of motions, it is of interest to study the influence of the Gilbert *precession* damping parameter $\alpha$ on the *nutation* resonance linewidth. The result is presented in Fig. 3 and is valid for ferromagnets with vanishing anisotropy. One sees that the dependence of $\Delta\Omega_N$ on $\alpha$ shows a minimum that becomes more expressive with increasing bias magnetic field. In other words, the linewidth is parametrized by the magnitude of field. This non-trivial behavior of linewidth relates with the nature of this oscillatory system, which performs two coupled motions.

To elucidate the non-trivial behavior, one can consider the susceptibility (9) in the same way as it is usually performed for the forced harmonic oscillator with damping [40]. For this oscillator, the linewidth can be directly calculated from the denominator of the response expression once the driving frequency is equal to eigenfrequency. In the investigated case of magnetization with inertia, the response expression is (9) with denominators (10) and (11) written as

$$D_\pm = \alpha^2 \tau^2 \omega^4 \pm 2\alpha\tau\omega^3 + \left(\alpha^2 - 2\alpha\tau\omega_H + 1\right)\omega^2 \mp 2\omega_H\omega + \omega_H^2. \quad (26)$$

Since the applied magnetic field is included in this expression as $\omega_H = |\gamma| H_0$, the linewidth depends on the field.

The obtained result can be generalized to a finite sample with magnetocrystalline anisotropy with method of effective demagnetizing factors [41,42]. In this case the bias magnetic field $\mathbf{H}_0$ denotes an external field and in the final expressions this field should be replaced by $\mathbf{H}_{i0} = \mathbf{H}_0 - \left(\hat{N}_a + \hat{N}_d\right)\mathbf{M}_0$, where $\hat{N}_a$ is the anisotropy demagnetizing tensor and $\hat{N}_d$ is the shape demagnetizing tensor.

## V. CONCLUSION

In summary, we derived a general analytical expression for the linewidth and frequency of nutation resonance in ferromagnets, depending on magnetization, the Gilbert damping, the inertial relaxation time and applied magnetic field. We show the nutation linewidth can be tuned by the applied magnetic field, and this tunability breaks the direct relation between losses and the linewidth. This for example leads to the appearance of a minimum in the nutation resonance linewidth for the damping parameter around $\alpha = 0.15$. The obtained results are valid for ferromagnets with vanishing anisotropy.

## ACKNOWLEDGEMENTS

We thank Benjamin Zingsem for helpful discussions. In part funded by Research Grant No. 075-15-2019-1886 from the Government of the Russian Federation, the Deutsche Forschungsgemeinschaft (DFG, German Research Foundation) − projects 405553726 (CRC/TRR 270), and 392402498 (SE 2853/1-1).